\documentclass[preprint2]{aastex}  

\def\etal{et al.\ }

\slugcomment{Submitted to the Astrophysical Journal}

\shorttitle{Weak Lensing from Space II}
\shortauthors{Richard Massey \etal}

\begin{document}

\title{Weak Lensing from Space II: \\ Dark Matter Mapping}

\author{Richard Massey}
\affil{Institute of Astronomy, Madingley Road, Cambridge CB3 OHA, U.K.}
\email{rjm@ast.cam.ac.uk}

\author{Alexandre Refregier\altaffilmark{1,2}}
\affil{Service d'Astrophysique, CEA Saclay,
        91191 Gif sur Yvette, France}

\author{Jason Rhodes\altaffilmark{3,4}}
\affil{California Institute of Technology, 1201 E. California Blvd,
        Pasadena, CA 91125}
\and

\author{Justin Albert$^{2}$, 
        David Bacon$^{5}$, 
        Gary Bernstein$^{6}$, 
        Richard Ellis$^{2}$, \\ 
        Bhuvnesh Jain$^{6}$, 
        Tim McKay$^{7}$, 
        Saul Perlmutter$^{8}$ \& 
        Andy Taylor$^{5}$}

\altaffiltext{1}{Institute of Astronomy, Madingley Road, Cambridge CB3 OHA,
         U.K.}
\altaffiltext{2}{California Institute of Technology, 1201 E. California Blvd,
         Pasadena, CA 91125}
\altaffiltext{3}{Laboratory for Astronomy \& Solar Physics, Code 681,
         Goddard Space Flight Center, Greenbelt, MD 20771}
\altaffiltext{4}{NASA/NRC Research Associate}
\altaffiltext{5}{Institute for Astronomy, Blackford Hill, Edinburgh EH9
         3HJ, U.K.}
\altaffiltext{6}{Department of Physics \& Astronomy, Univ. of Pennsylvania,
         209 S. 33$^{\rm rd}$ Street, Philadelphia, PA 19104}
\altaffiltext{7}{Department of Astronomy, University of Michigan, Ann
         Arbor, MI 48109}
\altaffiltext{8}{Lawrence Berkeley National Laboratory, 1 Cyclotron Road,
         Berkeley, CA 94720}

\begin{abstract}

We study the accuracy with which weak lensing measurements could be made from a
future space-based survey, predicting the subsequent precisions of 3-dimensional
dark matter maps, projected 2-dimensional dark matter maps, and mass-selected
cluster catalogues. As a baseline, we use the instrumental specifications of the
{\it Supernova/Acceleration Probe} (SNAP) satellite. We first compute its
sensitivity to weak lensing shear as a function of survey depth. Our predictions
are based on detailed image simulations created using `shapelets', a complete
and orthogonal parameterization of galaxy morphologies. We incorporate a
realistic redshift distribution of source galaxies, and calculate the average
precision of photometric redshift recovery using the SNAP filter set to be
$\Delta z=0.034$. The high density of background galaxies resolved in a wide
space-based survey allows projected dark matter maps with a rms sensitivity of
3\% shear in 1 arcmin$^2$ cells. This will be further improved using a proposed
deep space-based survey, which will be able to detect isolated clusters using a
3D lensing inversion techniques with a 1$\sigma$ mass sensitivity of
approximately $10^{13} M_\odot$ at $z=0.25$. Weak lensing measurements from
space will thus be able to capture non-Gaussian features arising from
gravitational instability and map out dark matter in the universe with
unprecedented resolution.

\end{abstract}

\keywords{dark matter --- gravitational lensing --- large-scale structure of
universe --- space vehicles.}

\section{Introduction} \label{intro}

Weak gravitational lensing has now been established as a powerful technique to
directly measure the large-scale mass distribution in the universe (for
reviews, see Mellier 1999; Bartelmann \& Schneider 2001; Refregier 2003).
Several groups have measured the coherent distortion of background galaxy
shapes around known galaxy clusters ({\it e.g.}\ Joffre \etal 2000; Dahle \etal
2002) and also statistically in the field ({\it e.g.}\ van Waerbeke \etal 2001;
Bacon \etal 2002; Hoekstra \etal 2002; Jarvis \etal 2003). Ever-growing surveys
using ground-based telescopes are beginning to yield useful constraints on
cosmological parameters (Bacon \etal 2003; Brown \etal 2002; Hoekstra, Yee, \&
Gladders 2002; van Waerbeke \etal 2002). The first two clusters selected purely
by weak lensing mass have now been found and spectroscopically confirmed by
Wittman \etal (2001, 2003).

Weak lensing is of such great interest for cosmology because it is directly
sensitive to mass. Other observations have traditionally been limited to
measuring the distribution of light and linked to theory via complications like
the mass-temperature relation for $x$-ray selected clusters (Pierpaoli, Scott
\& White 2001; Viana, Nicholl \& Liddle 2002; Huterer \& White 2003) or the
ubiquitous problem of bias (Weinberg \etal 2003). Weak lensing measurements
first avoid these problems, then have even been used to calibrate other
techniques (Huterer \& White 2003; Gray \etal 2002; Hoekstra \etal 2002b; Smith
\etal 2003). The high resolution, galaxy number density and stable image
quality available from space-based weak lensing data will allow maps of the
projected distribution of dark matter to be reconstructed at unprecedented
resolution. The mass power spectrum can be sliced into multiple redshift bins
using photometric redshifts: providing a long lever arm for constraints on the
evolution of cosmological parameters. Even three-dimensional mass maps,
marginally feasible from the ground (Bacon \& Taylor 2002), are likely to be
sensitive to overdensities as small as galaxy groups from space.

Mass-selected cluster catalogs can also be extracted from such maps (Weinberg
\& Kamionkowski 2002, Hoekstra 2002). Cluster counts, and the quantitative
study of high-sigma density perturbations or higher order shear correlation
functions (Bernardeau, van Waerbeke \& Mellier 1997; Cooray, Hu \&
Miralda-Escud\'{e} 2000; Munshi \& Jain 2001; Schneider \& Lombardi 2003) are
one of the most promising routes to breaking degeneracies in the estimation of
cosmological parameters including $\Omega_{\rm m}$ and $w$, the dark energy
equation of state parameter. Furthermore, studying well-resolved groups and
clusters individually, rather than statistically, will lead to a better
understanding of astrophysical phenomena, the nature of dark matter and the
growth of structure under the gravitational instability paradigm ({\it
e.g.}\ Dahle \etal 2003).

In this paper, we predict the general sensitivity to weak lensing of a
space-based wide field imaging telescope, taking as a baseline the
specifications of the proposed {\it Supernova/Acceleration Probe} (SNAP)
satellite. Instrument characteristics, including the PSF, ellipticity patterns
and image stability have been studied by Rhodes \etal (2003; paper I). In
$\S\ref{sims}$ we introduce detailed simulated images that have been developed
using shapelets, an orthogonal parameterization of galaxy shapes (Massey
\etal 2003, Refregier 2003). The simulated images contain realistic populations
and morphologies of galaxies as will be seen from space, modelled from those in
the Hubble Deep Fields (HDFs; Williams \etal 1996, 1998). These
shapelet-galaxies can be artificially sheared to simulate gravitational
lensing. The subsequent recovery accuracy of the known input shear is discussed
in $\S\ref{sens}$. We discuss the accuracy of SNAP photometric redshifts in
$\S\ref{photoz}$. These two measurements are combined to predict the accuracy
of projected dark matter maps, 3-dimensional dark matter maps and mass-selected
cluster catalogues in $\S\ref{maps}$. We draw conclusions in $\S\ref{conc}$.
Our results are used the predict the accuracy of cosmological parameter
constraints in Refregier \etal (2003; paper III).

\section{Image Simulations} \label{sims}

\begin{figure*}[htbp]
 \epsscale{2.1}
 \begin{center}
  \plottwo{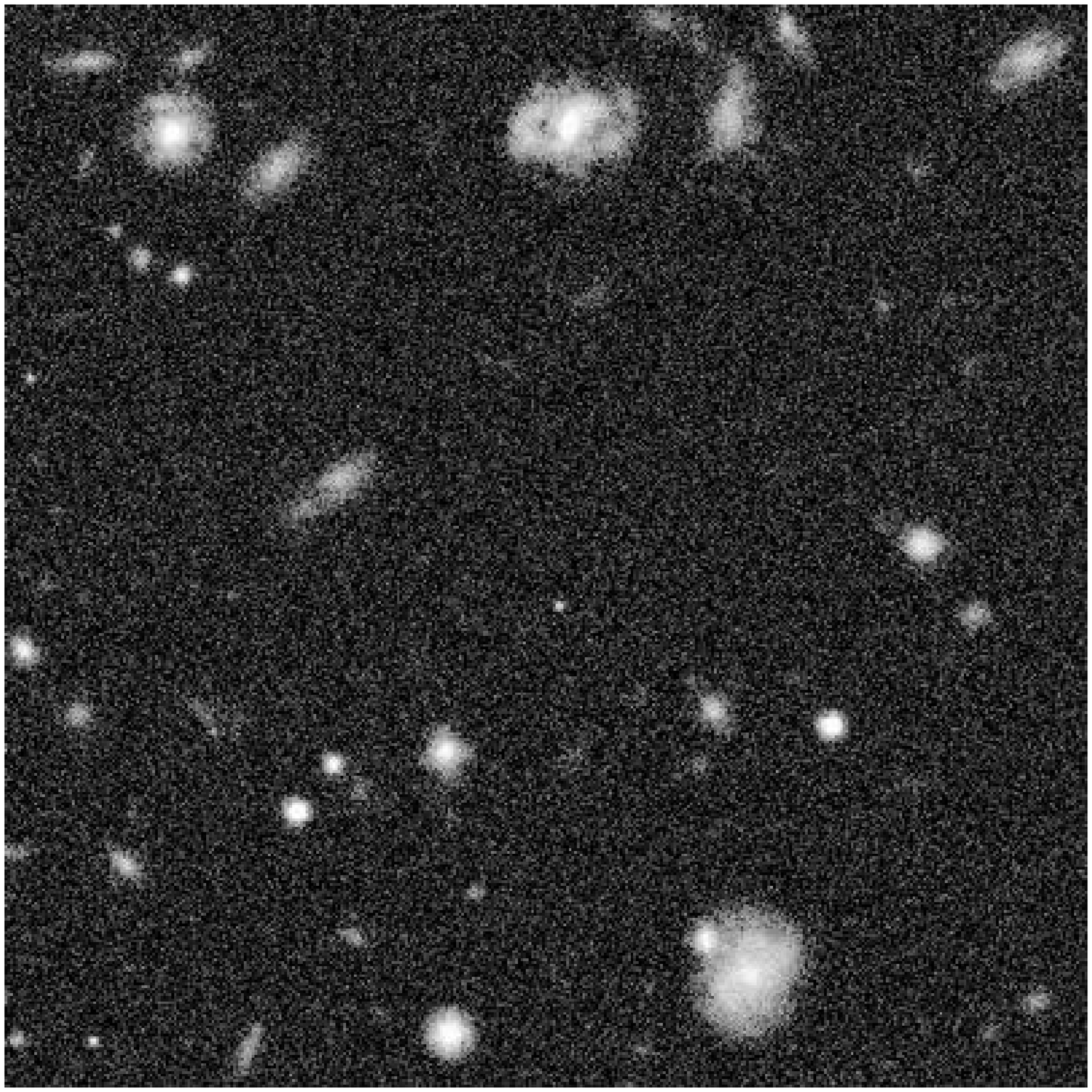}{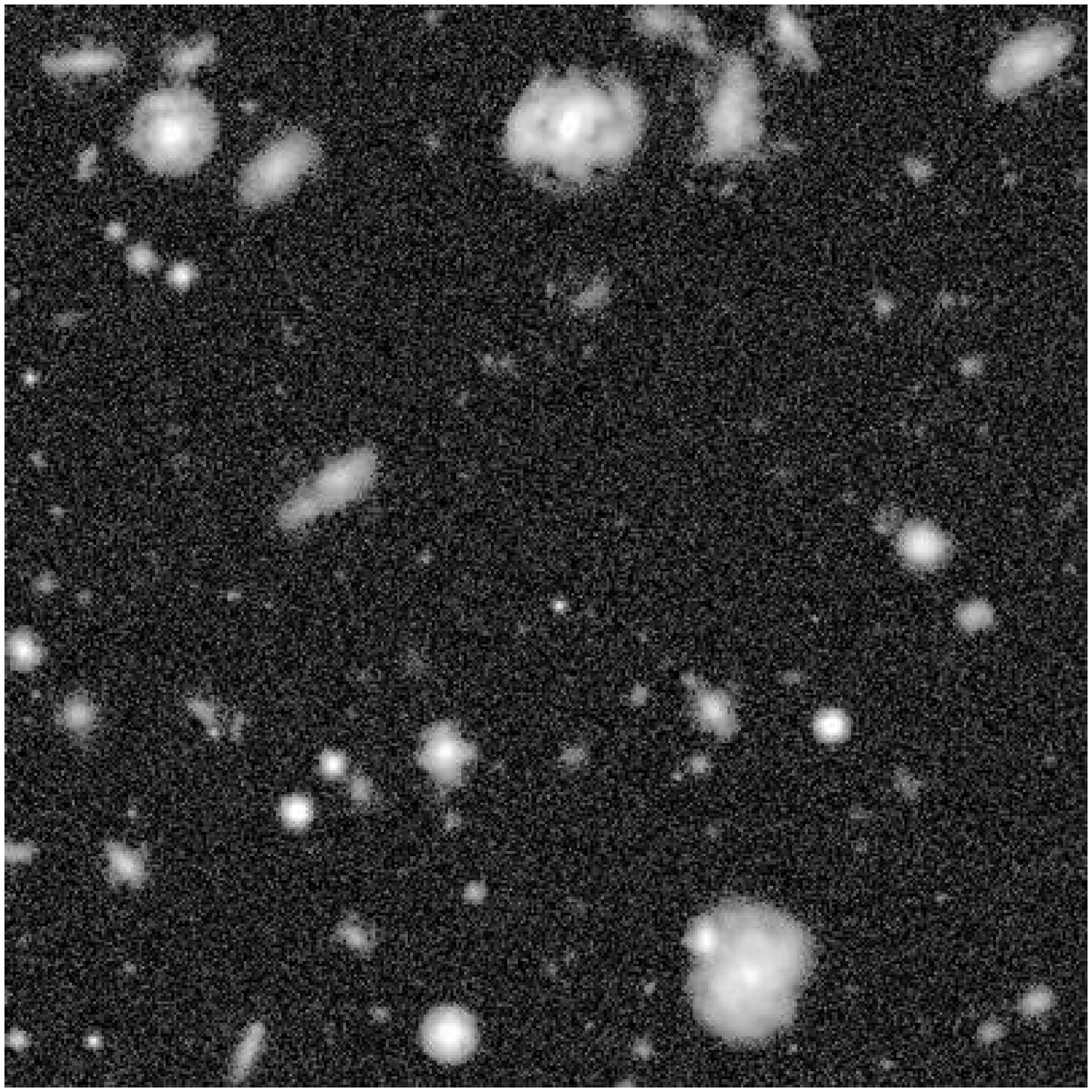}
 \end{center}
 \epsscale{1}
 \caption{\small 30$\arcsec\times$ 30$\arcsec$ portions of simulated SNAP $I$-band
 images, using the PSF shown in figure \ref{fig:psf}. {\it Left panel}: to the
 depth of the proposed SNAP wide survey. {\it Right panel}: to the
 depth of the Hubble Deep Fields. The SNAP deep survey will
 be some 2 magnitudes deeper than the latter, but further real data
 from the ACS on HST are needed to simulate this depth accurately.}
 \label{fig:images}
\end{figure*}

In this section, we describe our method for simulating realistic images, aiming
to closely resemble images observed with a space-based telescope. These
simulations are part of a full pipeline which allows us to propagate the
effects of perturbations in the instrument design onto shear statistics and
cosmological parameters. Example simulated images are shown in 
figure~$\ref{fig:images}$.


\subsection{Procedure} \label{makesims}

The shapelet formalism (Refregier 2003; Refregier \& Bacon 2003; summarized in
\S\ref{shapelets}) has been used to model all the galaxies in the HDFs. Using
just a few numbers, this parameterization captures the detailed morphology of
the galaxies, including spiral arms, arbitrary radial profiles and irregular
substructure. The parameters for each galaxy are stored in a multidimensional
parameter space. This is then randomly re-sampled, to simulate new and unique
galaxies with realistic properties as compared to those in the original HDFs.
A detailed description of the simulation procedure and performance can be
found in Massey \etal (2003).

The simulated images are built up with galaxies of all types (spiral,
elliptical and irregular) in their observed proportions, with realistic number
counts and a size distribution reproducing that in the HDF. Their morphology
distribution as a function of magnitude also reproduces that in the HDF. Most
importantly, all of these objects possess a precisely known shape, magnitude,
size and shear. The amount of shear can be adjusted in shapelet space as an
input parameter.

Observational effects including PSF convolution, pixelization, noise and
detector throughput are then incorporated in the simulations. In \S\ref{snap}
we describe the engineering specifications we have used to emulate the
performance of the SNAP satellite. In \S\ref{sens} we then attempt to recover
the known input shear from these realistic, noisy images using existing (and
independent) shear measurement methods.

\subsection{Shapelets} \label{shapelets}

Here we briefly describe the idea of shapelets, which is at the core of our
image simulation package. More comprehensive details are available in Refregier
(2003), Refregier \& Bacon (2003), and Massey \etal (2003). Shapelets are an
orthonormal basis set of 2D Gauss-Hermite functions. They can be used to model
any localized object by building up its image as a series of successive basis
functions, each weighted by a ``shapelet coefficient'', rather like a Fourier
or wavelet transform. Each polar basis state and shapelet coefficient can be
identified by two integers: $n\geq 0$ describing the number of radial
oscillations, and $m\in\{-n,n\}$ the azimuthal oscillations, or rotational
degrees of symmetry. The basis is complete when the series is summed to
infinity, but it is truncated in practice at a finite $n_{max}$. This offers
image compression because an object is typically well-modelled using only a few
shapelet coefficients.

Conveniently, the shapelet coefficients are Gaussian-weighted multipole moments
(with the rms width of the Gaussian known as the shapelet scale size $\beta$),
as commonly used in various astronomical applications. The $n=2$ states are
thus Gaussian-weighted quadrupole moments, the $n=4$ states octopole moments,
{\it etc}. Shapelet basis functions also happen to be eigenstates of the 2D
Quantum Harmonic Oscillator, with $n$ and $m$ corresponding respectively to
energy and angular momentum quantum numbers. This analogy suggests a
well-developed formalism. For instance, shears and dilations can be represented
analytically as $\hat{a}$ or $\hat{a}^\dag$ ladder operators (Refregier 2003);
and PSF convolutions as a trivial bra-ket matrix operation (Refregier \& Bacon
2003).

Massey \etal (2003) demonstrate how HDF galaxies can be represented as
shapelets and then transformed by slight adjustments of their shapelet
coefficients into new shapes. This process produces genuinely new but realistic
galaxies, as proved by the similar distributions in HDF and simulated data of
commonly used diagnostics from {\sc SExtractor} (Bertin \& Arnouts 1996) and
galaxy morphology packages ({\it e.g.} Conselice, Bershady \& Jangren 2000).

\subsection{SNAP Simulations} \label{snap}

For this work, our image simulations have been tuned to the instrument and
specifications of the proposed SNAP mission (paper I; Aldering \etal 2002; Kim
\etal 2002; Lampton \etal 2002a, 2002b; Perlmutter \etal 2002). The SNAP
strategy includes a wide, 300 square degree survey (with 4$\times$500s
exposures reaching a depth of AB 27.7 in R for a point source at $5\sigma$),
and a deep, 15 square degree survey (120$\times$4$\times$300s to AB 30.2). For
an exponential disc galaxy with FWHM=0.12", these limits become 26.6 and 28.9
respectively.

\begin{figure}[htbp]
 \plotone{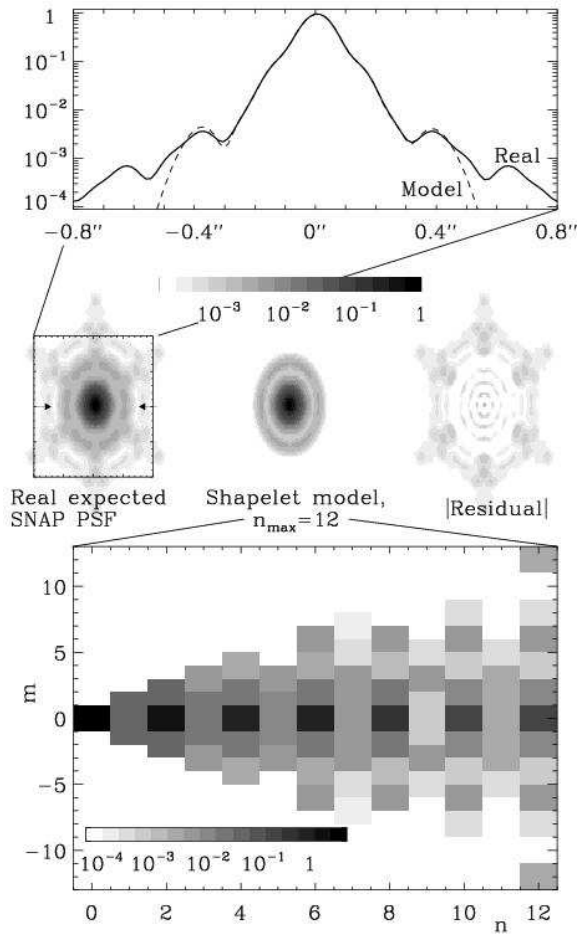}
 \caption{\small Shapelet decomposition of the proposed SNAP PSF. {\it Top
 panel}: a horizontal slice through the center of the real (solid line)
 and shapelet-reconstructed (dashed line) PSF.
 The middle panels shows, in 2-dimensions, the real PSF, its recovery using
 shapelets and the residual difference between the two, from left to
 right respectively. {\it Bottom panel}: the moduli of the corresponding
 polar shapelet coefficients with order up to $n_{max}=12$. Note that all
 intensity scales are logarithmic. The circular ($m=0$) core is modelled to an
 accuracy of about $10^{-3}$ and the beginnings of six-fold symmetric
 structure is seen as power in the $m=\pm 6,\pm 12$ shapelet coefficients.
 \label{fig:psf}}
\end{figure}

The predicted SNAP PSF at the middle of the illuminated region of the focal
plane is illustrated in figure $\ref{fig:psf}$. Following the analysis of paper
I, this was obtained for the current satellite design, using raytracing,
aperture diffraction and CCD diffusion. In this paper we also illustrate the
decomposition of the SNAP PSF into shapelets. As shown on the top panel of
figure $\ref{fig:psf}$, our model includes the second diffraction ring and is
accurate to nearly one part in 10$^3$. It does not include much of the extended
low-level diffraction spikes, which we ignore. Convolution with this residual
PSF pattern adds less than 0.7\% to the ellipticity of any exponential disc
galaxy that passes the size cut into the lensing catalog (see
$\S\ref{method}$). Given the further factor of $G^{-1}$ in
equation~[$\ref{eq:g-e}$], to convert ellipticity into shear, this residual
thus has a negligible impact upon shear measurement within the accuracy of the
current methods.

Simulated images used to calibrate the shear measurement method (see
\S\ref{method}) were first sheared and then convolved with the full SNAP PSF
shown in figure $\ref{fig:psf}$. For this application, it is essential that the
shearing is applied before the smearing, just as occurs in the real universe.
Shear measurement methods have been designed to correct for precisely this
sequence of events. However, our simulated galaxies were modelled on real HDF
objects which had already been naturally convolved with the WFPC2 PSF when the
HDF images were taken. Consequently, our simulated objects in \S\ref{method}
exhibit smoothing from both a circularised WFPC2 PSF, (plus shearing), plus a
SNAP PSF. This double PSF artificially reduces the rms ellipticity of galaxies
by approximately $\sim$2\% and increases the size of a point source by 22\%.
One should note that the first PSF convolution occurs, and the galaxy
orientations are randomized, all before shearing. This effect therefore
corresponds to a small alteration in the intrinsic shape distribution of
galaxies but does not bias the shear measurement (see discussion in Massey \etal 2003).

Simulated images used to predict the lensing efficiency as a function of
exposure time (see \S\ref{results}) were produced differently. For these, we
needed to ensure realistic size distributions and number counts in the
simulations. The galaxies had no artificial shear added: they just have a
scatter of ellipticities due to their own intrinsic shapes. We convolved these
galaxies by the PSF difference between the HST and SNAP. This is obtained by
deconvolving the WFPC2 PSF from the SNAP PSF model, in shapelet space.
Smoothing an object with this smaller kernel is enough to convert it from an
observation with HST to one with SNAP, although without inputting shear.

\begin{figure}[htb]
 \plotone{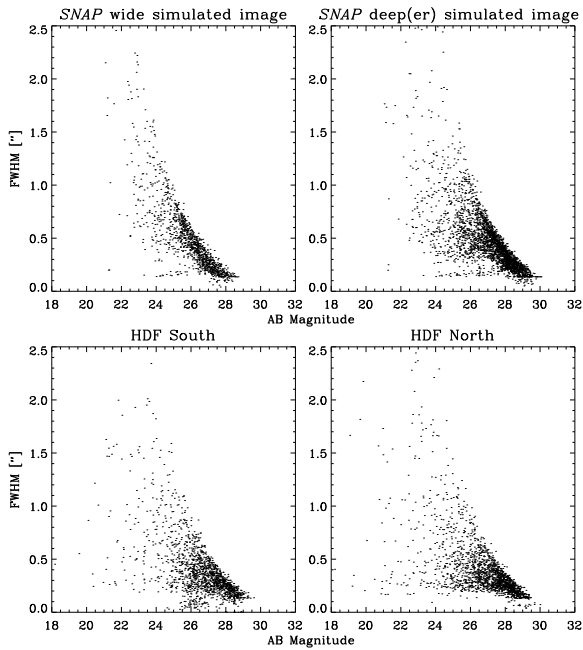}
 \caption{\small Size vs magnitude as determined by {\sc SExtractor} with a $S/N$
 cutoff at $\nu=1.5$. Top panels are for simulated SNAP $I$ band images of
 the same size as the Hubble Deep Field. For reference, the bottom panels
 are of the HDFs themselves using the same {\sc SExtractor} parameters.}
 \label{fig:sizemag}
\end{figure}

Example simulated images are shown in figure $\ref{fig:images}$ for
the wide SNAP survey (left panel) and to the depth of the HDF (right
panel). They include a noise model consisting of both photon counting
error and a Gaussian background. These compare well with real deep HST
images (see Massey \etal 2002). The SNAP deep fields will be about 2
magnitudes deeper than the HDF. However, deeper surveys with the ACS
on board HST are awaited to accurately model galaxies at this depth.
Figure $\ref{fig:sizemag}$ shows the size-magnitude distribution of
the simulated images to both depths (top panels). Again, the
simulations reproduce the statistics of the real HDFs (bottom panels).

\subsection{Limitations of the Simulations} \label{simlims}

The SNAP wide survey strategy includes four dithered exposures at each
pointing. This will enable the removal of cosmic rays and, if necessary, the
simultaneous measurement of instrumental distortions. Because of the high orbit
and slow thermal cycle, instrument flexure and the PSF are expected to be very
stable (see paper I). It should therefore be possible to map internal
distortions and compensate for them even on small scales, using periodic
observations of stellar fields. Consequently, neither cosmic rays nor
astrometric distortions are added to the simulations.

The SNAP CCD pixels are 0.1$\arcsec$ in size and thus under-sample
the PSF. To compensate for this, the dithered exposures will be
stacked, as usual for HST images, using the  {\sc DRIZZLE}
algorithm (Fruchter \& Hook 2002). Alternatively, galaxy shapes
may be fitted simultaneously from several exposures. {\sc DRIZZLE}
recovers some resolution, and will be particularly effective for
the multiply-imaged SNAP deep survey, but has the side-effect of
aliasing the image and correlating the noise in adjacent pixels.
We have not yet included this entire pipeline in the simulations,
but merely implemented a smaller pixel scale and model background
noise that is higher in each pixel (although uncorrelated).
Following the example of the Hubble Deep Field final data
reduction, we choose 0.04$\arcsec$ pixels. Unfortunately, the
detection and shape measurement of very faint galaxies is
sensitive to the precise noise properties of an image. Because of
these instabilities, our simulated images are only reliable down
to approximately $I\simeq29.5$ (see Massey \etal 2003). This is
just below the magnitude cut applied by our shear measurement
method at $I=29.1$. A further investigation will include full use
of {\sc DRIZZLE} and more detailed noise models. This will also
address the issue of pointing accuracy, and consider the
consequences of `dead zones' around the edges of the pixels which
house the CCD electronics and are therefore unresponsive to light.

The image simulations are based upon the galaxies in the HDF, which is itself
a special region of space selected to contain no large or bright objects. As a
result, our simulations do not yet include these either. The source catalog is
being expanded as GOODS ACS data becomes publicly available.

The image simulations are currently mono\-chromatic, in the HST $F814W$
(hereafter $I$) filter. Since gravitational lensing is achromatic, shear
measurement can be performed in any band: indeed, all tested shear measurement
methods so far use only one color at a time. $I$ or $R$ bands are typically
chosen for shear measurement because of the increased galaxy number density,
advanced detector technology, and small PSF at these wavelengths. Surveys like
COMBO17 (Brown \etal 2003), and VIRMOS/Descartes (van Waerbeke \etal 2002) are leading a trend to use additional multicolor photometry to
provide photometric redshifts of the source galaxy population. The SNAP surveys
will be simultaneously observed in 9 bands: 6 optical colors spanning roughly
$B\rightarrow I$, plus $J$, $H'$ and $K$ (the near IR filters are twice as
large and receive double the total exposure times given in $\S\ref{makesims}$;
see paper I). We have not simulated this multicolor data, but it will
inevitably raise the S/N of shear estimation for every source galaxy. At a
minimum, image coaddition or simultaneous fits to shapes in several colors will
increase the effective exposure time. Something more ambitious, like shifting
to the rest-frame $R$ or the rotating disc dis-alignment suggested by Blain
(2002), might even reduce systematic measurement biases. Further work is needed
in cosmic shear methodology to investigate the optimal use of multicolor data.
However, it can already be said that our current monochromatic approach will
yield a conservative estimate of the lensing sensitivity expected from future
analyses.

\section{Weak Lensing sensitivity} \label{sens}

In this section, we determine the accuracy with which it is possible to recover
the input shear from the noisy image simulations. The formalism of shapelets
can be used to form an accurate shear measurement (Refregier \& Bacon, 2003).
However, since the images themselves were created using shapelets, we choose
here to be conservative and use a slightly older but independent method
developed by Rhodes, Refregier \& Groth (2000; RRG).

\subsection{Advantages of space} \label{advspace}

We first discuss the advantages specific to weak lensing measurement that are
provided by observations from space. The figure of merit for any lensing survey
needs to include more than the {\it \'{e}tendue}, a product of the survey area
and the flux gathering power of a telescope (Tyson \etal 2002, Kaiser \etal
2002). It must also account for the finite PSF size, the size-magnitude
distribution of background galaxies, and systematics ({\it e.g.}\ due to the
atmosphere or telescope optics). Shear sensitivity is raised for a spacecraft
over a ground-based telescope for the additional reasons listed below.

\begin{itemize}

\item More objects have measurable shapes. Although not as much
sky area will be surveyed as by proposed ground based surveys such
as MEGACAM  (Boulade \etal 2000), VISTA ({\tt
http://www.vista.ac.uk}), or LSST ({\tt http://www.lsst.org}), the
number density of resolved objects is an order of magnitude higher
from space (compare figure \ref{fig:texp} with Bacon \etal 2001).
Such an increase in $S/N$ per unit area will enable the mapping of
projected dark matter maps with adequate resolution for a direct
comparison to redshift surveys ($\S\ref{2dmap}$) and the
generation of a mass-selected cluster catalog ({\it e.g.}\
Weinberg \& Kamionkowski 2002, Hoekstra 2002). Quantitative study
of high-sigma mass fluctuations is one of the most promising
methods to break degeneracies in cosmological parameter
estimation, particularly constraining $\Omega_{\rm m}$ ({\it
e.g.}\ van Waerbeke \& Mellier 1997; Cooray, Hu \&
Miralda-Escud\'{e} 2000; Munshi \& Jain 2001; Schneider 2002).
Furthermore, studying well-resolved groups and clusters
individually, rather than statistically, will lead to a better
understanding of astrophysical phenomena such as biasing or the
mass-temperature relation (Weinberg \etal 2002; Huterer \& White;
2003 Smith \etal 2003).

\item The shape of individual galaxies are more precisely measured. The SNAP
PSF is small (0.13$\arcsec$ FWHM assuming 4$\mu$m CCD diffusion). It is more
isotropic and, importantly, more stable than even the HST PSF (see paper I).
This enables shape measurement to be more reliable, or possible at all, for
small, distant galaxies. The stable photometry from the 3-day orbit may even
permit the use of weak lensing magnification as well as shear information (see
{\it e.g.} Jain 2002; 2003). Whether directly measured or inferred from shear,
this in turn is useful to correct for the effect of lensing on the distance
moduli to the SNAP supernov\ae\ (Dalal \etal 2003; Perlmutter \etal 2002).

\item Galaxy redshifts are known accurately and to a greater depth. SNAP's
stable 9-band optical and NIR imaging is ideal to produce exquisite photometric
redshifts for almost all galaxies at $z\lesssim3$ detected at 5$\sigma$ in the
$I$-band (see $\S\ref{photoz}$). This should be compared to the $\sim$38\%
completeness of photo-$z$s possible from the ground in the COMBO-17 data with a
similar cut and a median redshift of $R\simeq24$ (Brown \etal 2002).
This allows a good estimation of the redshift distribution of source galaxies,
the uncertainty in which is a major contribution to the error budget in current
lensing surveys. Projected 2D power spectra and maps can be drawn in several
redshift slices, using redshift tomography.  More ambitiously, cluster catalogs
and dark matter maps can be constructed directly in 3D ($\S\ref{3dmap}$),
enabling the 3D correlation of mass and light and the tracing of the growth of
mass structures.

\item Galaxies are farther away. Distant objects, too faint and too small to
be seen from the ground, are measurable from space. The evolution of
structures can thus be traced from earlier epochs, giving a better handle on
cosmological parameters (see paper III). Furthermore, recent numerical
simulations (Jing 2002; Hui \& Zhang 2002) suggest that intrinsic galaxy
alignments impact lensing surveys to a greater depth in redshift than
previously assumed. If this is confirmed, intrinsic alignments will mimic and
bias cosmic shear signal in all but the deepest surveys, where the galaxies
are farther apart in real space. Using 3D positions of galaxies from SNAP
photo-$z$s, it will be possible to isolate close galaxy pairs and to measure
their alignments, or to optimally down-weight close pairs thus reducing their
impact (King \& Schneider 2002; Heavens \& Heymans 2002).

\end{itemize}

\subsection{Shear measurement method} \label{method}

The advantages of space-based data described above will provide limited gains
without an equally precise and robust shape measurement method. The
now-standard weak lensing method for ground-based data was introduced by
Kaiser, Squires \& Broadhurst (1995; KSB). KSB forms shear estimators from
quadrupole and octopole moments of an object's flux. Modern techniques are
being developed to incorporate higher order shape moments or Bayesian
statistics to raise the sensitivity to shear. These methods include shapelets
(Refregier \& Bacon, 2003) and others by Bernstein \& Jarvis (2001); Bridle
\etal (2003); Kaiser (2000). However, since the simulations themselves were
created using shapelets, we choose here to be conservative and use the
independent method developed by Rhodes, Refregier \& Groth (2000; RRG). This is
related to KSB, but optimized for use with space-based data. It has already
been used extensively on HST images (Rhodes \etal 2001; Refregier \etal 2002)
and is therefore appropriate for our current purposes.

Following KSB, RRG measures a galaxy's two-component ellipticity
$\varepsilon_{i}$ from the Gaussian weighted quadrupole moments of its surface
brightness $I({\mathbf \theta})$,

\begin{equation}
\label{eq:e}
 \varepsilon_{i} \equiv
  \frac{\{J_{11}-J_{22},~2J_{12}\}}{J_{11}+J_{22}}
\end{equation}
where
\begin{equation}
\label{eq:Jij}
 J_{ij}       \equiv
  \frac{\int d^{2}\theta~\theta_{i}~\theta_{j}~
        w({\mathbf \theta})~I({\mathbf \theta})}
       {\int d^{2}\theta~w({\mathbf\theta})~I({\mathbf\theta})}~,
\end{equation}

\noindent and $w({\mathbf \theta})$ is a Gaussian of width adjusted to match
the galaxy size. The unweighted PSF moments are measured from a (simulated)
starfield and RRG corrects the galaxy ellipticities to first order for PSF
smearing. Occasional unphysical ellipticities, $|\varepsilon|>2$, are excluded,
along with galaxies fainter than AB 26.5 (for the SNAP wide survey) or AB 29.1
(for the SNAP deep survey) and with sizes

\begin{equation}
  \label{eq:size}
  R \equiv \sqrt{\frac{1}{2}\left(J_{11}+J_{22}\right)} \le 1.7\mathrm{pixels}.
\end{equation}

\noindent Note that $R$ is an rms size measure rather than a FWHM, and that
this procedure does indeed select only resolved objects. The locations of these
cuts have been chosen to yield reasonably stable results; the effect of moving
the size cut is discussed further in section~\S\ref{size_cut}. 

RRG finally provides the shear susceptibility conversion factor, $G$, to
generate unbiased shear estimators $\hat{\gamma}_{i}$ for an ensemble of
objects, given by

\begin{equation}
  \label{eq:g-e}
  \hat{\gamma}_{i} = \frac{ \langle \varepsilon_{i} \rangle}{G}~,
\end{equation}

\noindent where $G$ depends upon the fourth order moments $J_{ijkl}$ of a
galaxy population, defined similarly to equation~[\ref{eq:Jij}]. In our
simulated SNAP images, $G$ is of order 1.6.

\begin{figure}[htbp]
 \plotone{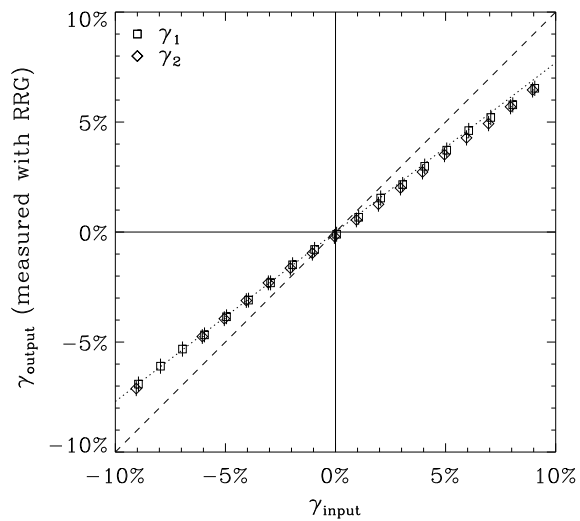}
 \caption{\small The applied shear $\gamma_{in}$ in the shapelet simulated images
 vs its recovery $\gamma_{out}$ using an independent measurement method
 (Rhodes, Refregier \& Groth 2000). The image used is one $7.5$
 arcmin$^{2}$ realization of the SNAP deep survey shown in figure
 $\ref{fig:images}$. The recovery is linear, but the slope of the fitted
 line (dotted) is flatter than that expected (dashed line).}
 \label{fig:ginout}
\end{figure}

This shear measurement method and the simulations are tested in figure
$\ref{fig:ginout}$. An artificial shear is applied uniformly upon all objects
in a $7.5$ arcmin$^{2}$ simulated image, in the $\gamma_2=0$ and $\gamma_1=0$
directions, before convolution with the SNAP PSF. Using RRG, we correct for the
PSF smearing and recover the input shear. As can be seen in figure
$\ref{fig:ginout}$, the recovery is linear, but the slope (see dotted line) is
underestimated (dashed line). This inconsistency probably has two origins:
inaccuracy of the image simulations and instabilities in the shear measurement
method. The latter may be removed with future techniques. For the purposes of
this paper, we follow the procedure adopted by Bacon \etal (2001), where
a similar bias was observed in the KSB method. We apply a linear correction
factor to the measured shears and to their errors. This factor is
$(0.79\pm0.03)^{-1}$ at the depth of the HDF, and $(0.87\pm0.04)^{-1}$ for the
SNAP wide survey.

Even after this correction, there remains a small difference in the rms scatter
of galaxy ellipticities between the simulations and real Groth strip data
(Rhodes, Refregier \& Groth 2001). As shown in Massey \etal (2003), this
discrepancy is not detected with the standard shape measures of {\sc
SExtractor} (Bertin \& Arnouts 1996), however RRG proves to be a more sensitive
test. Perhaps because of the precise properties of the background noise, or
perhaps because the wings of simulated objects are truncated beyond the {\sc
SExtractor} isophotal cutoffs, $\sigma_e$ is observed by RRG to be lower in the
simulated images by another factor of $\sim0.8$. Work is in progress to
establish the origin of this effect. For the purposes of this paper, we simply
increase the error bars by this amount.

\subsection{Shear sensitivity of SNAP} \label{results}

Now that the image simulation and analysis pipeline is in place, we can measure
SNAP's sensitivity to shear. Trade-off studies are under way for several
alternative telescope designs, including the level of CCD charge diffusion, the
pixel size, the effect of {\sc DRIZZLE}ing, and the coefficient of thermal
expansion in the secondary struts, which may be the main cause of temporal
variation in the PSF (see paper I). Here we present the results of a study
which uses the baseline design specifications and time-averaged PSF of the SNAP
satellite. In this study the PSF used is the residual between the HST and SNAP
PSFs (see \S\ref{snap}), in order to keep the size distribution of galaxies
realistic for SNAP images.

\begin{figure}[htbp]
 \plotone{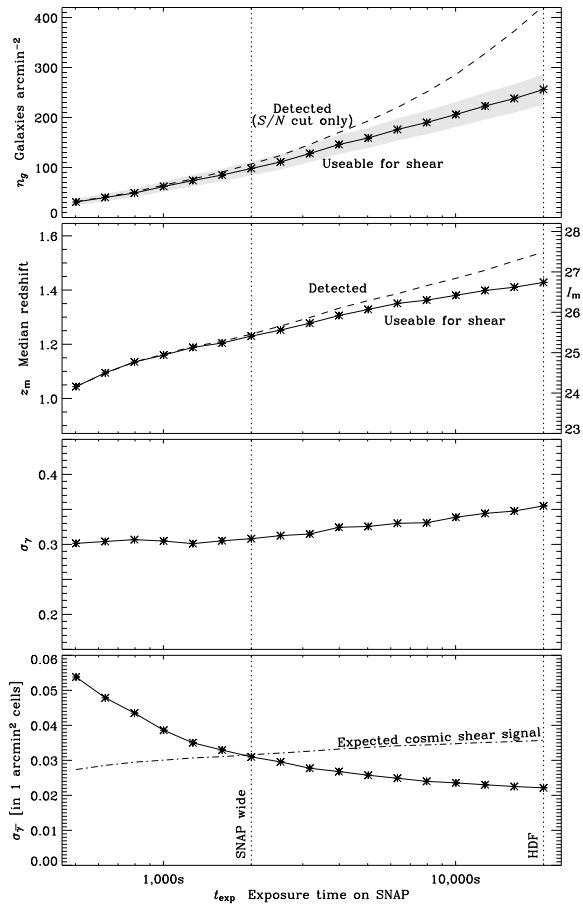}

 \caption{\small Shear sensitivity as a function of SNAP exposure time
 $t_{\rm exp}$. {\it Top panel}: the surface number density of all
 galaxies ($n_{g,{\rm tot}}$) detected by {\sc SExtractor} and of the
 subset ($n_{g}$) of these useable for weak lensing, {\it i.e.}\ having
 survived further cuts in size and ellipticity by RRG (see
 text). {\it Second panel}: the median $I$ band magnitude, $I_{\rm m}$,
 in the two subsets of the galaxy catalog, which has been interpreted
 as median redshift, $z_{\rm m}$, using equation~[\ref{eq:itoz}]. {\it Third
 panel}: the rms error $\sigma_{\gamma}=\langle
 |\gamma|^{2} \rangle^{1/2}$ per galaxy for measuring the shear $\gamma$,
 after PSF correction and shear calibration. {\it Bottom panel}: 
 the rms error $\sigma_{\overline{\gamma}}$ for measuring the mean
 shear $\overline{\gamma}$ in 1 arcmin$^{2}$ bins. The dot-dashed line
 shows an estimate of the expected rms shear in a $\Lambda$CDM universe.}

 \label{fig:texp}
\end{figure}

The top panel of figure $\ref{fig:texp}$ shows the surface number density
$n_{g}$ of galaxies in a survey of a given exposure time $t_{\rm exp}$ on SNAP.
The exposure times reflect a $\sim5\times$ overall improvement in instrument
throughput and detector efficiency over WFPC2 on HST (Lampton \etal 2002). The
dashed line shows the number density $n_{g,{\rm tot}}$ of all the galaxies
detected by {\sc SExtractor}, after a $S/N$ cut which is equivalent to $I<29.1$
at the depth of the HDF. As discussed in $\S\ref{method}$, galaxies which are
too faint, too small, or too elliptical are excluded from weak shear catalogs.
The solid line shows the number density $n_{g}$ of galaxies which are useable
for weak lensing following the magnitude, size and ellipticity cuts. The error
bars reflect the uncertainty in measuring number counts at low $S/N$ and an
estimated sample variance between the HDF-N and HDF-S.

An important cut in the weak lensing analysis is the size cut, which reduces the
detected galaxy sample by about 30\% at the depth of the HDF. This fraction is a
strong function of PSF size, and is thus much larger for ground based imaging.
As can be inferred from the top panel of figure $\ref{fig:texp}$, the SNAP wide
survey ($n_{g}\simeq 100$ galaxies arcmin$^{-2}$) will thus provide a dramatic
improvement over current ground-based surveys ($n_{g}\sim 25$ galaxies
arcmin$^{-2}$ are used by most groups; see {\it e.g.}\ Bacon \etal 2002). The
effect of moving the size cut is discussed further in section~\S\ref{size_cut}. 

The second panel of figure $\ref{fig:texp}$ shows the median magnitude, $I_{\rm
m}$, of the galaxy catalog before and after cuts in size and ellipticity by the
weak lensing analysis software. This has been converted to a median redshift,
$z_{\rm m}$, using equation~[\ref{eq:itoz}]. For the purposes of this plot, we
assume that this relationship is still valid even after the size cut.

The third panel of figure $\ref{fig:texp}$ shows the rms error
$\sigma_{\gamma}=\langle|\gamma|^2\rangle^{1/2}$ per galaxy for measuring the shear,
after the PSF correction and shear calibration. The slightly increasing error
at longer $t_{\rm exp}$ reflects the decreasing size of fainter galaxies, and
correspondingly less resolved information content available about their shapes.
To map the shear, the noise can be reduced by binning the galaxies into cells.
The rms noise of the shear $\overline{\gamma}$ averaged in a cell of solid
angle $A=1$ arcmin$^2$ is given by

\begin{equation}
\label{eq:sigma_gamma} \sigma_{\overline{\gamma}} \simeq
\frac{\sigma_{\gamma}}{\sqrt{n_{g} A}},
\end{equation}

\noindent and is plotted in the bottom panel of figure $\ref{fig:texp}$. The
wide and deep SNAP surveys will thus afford a $1\sigma$ sensitivity for the
shear of $\simeq3.0$\% and better than $2.2$\% on this scale, respectively. As
a comparison, the rms shear expected from lensing on this scale in a
$\Lambda$CDM model is approximately 3\% (assuming $\Omega_{\rm m}$=0.3,
$\Omega_\Lambda$=0.7, $\sigma_8$=0.9, $\Gamma$=0.21). This signal increases
with survey depth because the total lensing along a line of sight is
cumulative. The wide SNAP survey will thus be ideal to map the mass
fluctuations on scales of 1 arcmin$^2$, with an average $S/N$ of unity in each
cell. The recovery of simulated mass maps will be discussed in $\S\ref{maps}$.

Note that the shear sensitivities presented here are conservative estimates,
particularly for the deep SNAP survey. The image simulations extend so far only
to the depth of the HDF. Future shear measurement methodology will also be more
accurate and stable on any individual, resolved galaxy than the RRG method used
in this paper. Higher order shape statistics ({\it e.g.}\ shapelets) will be
used, as will simultaneous measurements in multiple colors and pre-selection of
early-type galaxy morphologies.

\subsection{Effect of size cut and pixel scale} \label{size_cut}

\begin{figure}[tb]
 \plotone{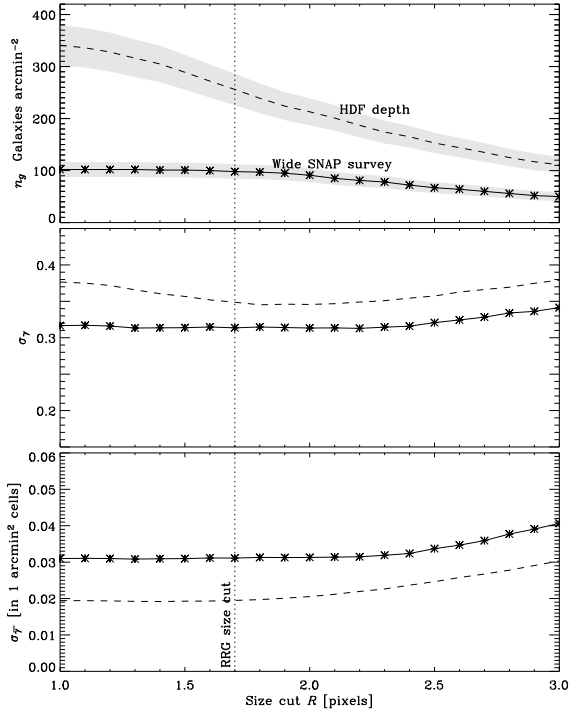}
 \caption{\small Shear sensitivity as a function of size cut $R$ in the RRG shape
 measurement method for the wide SNAP survey (solid line) and at the depth of
 the HDF (dashed line). The vertical dotted line shows the fiducial value
 adopted elsewhere in our analysis. {\it Top panel}: the surface number density of galaxies
 useable for weak lensing. {\it Middle panel}: the rms error
 $\sigma_{\gamma}=\langle |\gamma|^{2} \rangle^{1/2}$ per galaxy for measuring
 the shear $\gamma$, after PSF correction and shear calibration. {\it Bottom
 panel}: the rms error $\sigma_{\overline{\gamma}}$ for measuring the mean
 shear $\overline{\gamma}$ in 1 arcmin$^{2}$ bins.} 
 \label{fig:size_cut}
\end{figure}

Small, faint and highly elliptical objects are excluded from the final galaxy
catalog in the RRG shear measurement method. Of all these cuts, it is the size
cut that excludes the most objects. In an image at the depth of the HDF, about
30\% of detected galaxies are smaller than our adopted size cut at
$R=1.7$~pixels. The exact position of this cut has been determined empirically
to produce stable results, from experience with both HST data and our simulated
images. The quantitative effects of moving the size cut are demonstrated in
figure~\ref{fig:size_cut}. 

If the cut is moved to a larger size, fewer objects are allowed into the final
galaxy catalog, and the shear field is sampled in fewer locations.
Consequently, both dark matter maps and cosmic shear statistics become more
noisy. If smaller galaxies are included in the catalog, the shear field is
indeed better sampled, but the shape measurement error is worse on these
galaxies. The bottom panel of figure~\ref{fig:size_cut} shows that moving the
size cut to smaller objects has no net change in the precision of shear
recovery: adding noisy shear estimators to the catalog neither improves nor
worsens the measurement. A size cut at $R=1.7$~pixels is optimal at the depth
of the HDF and in the observing conditions modelled by our image simulations.
To simplify comparisons of galaxy number density, the same cut has been applied
to data at the depth of the SNAP wide survey. A different cut could have been
adopted, producing fewer galaxies but each with more accurate shear estimators:
the crucial figure $\sigma_{\hat\gamma}$ would not change. (This is especially
true in the SNAP wide survey because of the relative dearth of small galaxies).

As described in section~\ref{simlims}, we have assumed that an effective image
resolution of 0.04'' can be recovered for SNAP data by taking multiple,
dithered exposures, and either stacking them with the {\sc DRIZZLE} algorithm
or by fitting each galaxy's shape simultaneously in them all. The increase in
image resolution from these techniques is vital for cosmic shear measurements.
The number density of useable galaxies increases dramatically, and the
measurement of their shapes is improved. Were it not possible to apply {\sc
DRIZZLE} or to recover this resolution, the large pixel scale currently
proposed for SNAP would seriously impair shear measurement. A size cut at
$R=0.12\arcsec$ ($=3$ pixels in figure~\ref{fig:size_cut}) would roughly halve
the number density of useable sources and correspondingly reduce the
sensitivity to gravitational lensing.

\section{Photometric Redshift Accuracy} \label{photoz}

Gravitational lensing is achromatic, so shear measurement may be performed in
any color. As discussed in $\S\ref{simlims}$, current techniques measure galaxy
shapes in only one band at a time (usually $R$ or $I$ are chosen for their
steeper slope of number counts). However, gravitational lensing is also a
purely geometrical effect, and measurements are aided greatly by accurately
knowing the distances to sources. The latest surveys, and future high-precision
measurements will therefore require multiple colors for photometric redshift
(photo-$z$) estimation. Reliable photo-$z$s will not only remove current errors
due to uncertainty in the redshift distribution of background sources, but even
make possible an entirely 3D mass reconstruction, as demonstrated in
$\S\ref{3dmap}$, Taylor (2003a), Hu \& Keeton (2002), Bacon \& Taylor (2002)
and Jain \& Taylor (2003).

SNAP's thermally stable, 3-day long orbit is specifically designed for
excellent photometry on supernov\ae. Combining all 9 broad-band filters (6
optical, 3 NIR) will also provide an unprecedented level of photo-$z$ accuracy,
for all morphological types of galaxies over a large range of redshifts. In
this section, we simulate SNAP photometric data in order to determine this 
precision.

We have used the {\sc hyperz} code (Bolzonella, Miralles \& Pell\'{o}
2000) to generate the observed magnitudes of a realistic catalog
of galaxies following Lilly \etal (1995),
\begin{equation} \label{eq:dNdI}
\frac{{\mathrm d}N}{{\mathrm d}I}(I) \simeq  10^{0.35\times I}~,
\end{equation}
\noindent where $I$ is the $I$-band magnitude. The galaxies were
assigned a distribution of Spectral Energy Distribution (SED)
types similar to that in real data and containing ellipticals,
spirals and starburst galaxies. Redshifts were assigned at random,
and independently of spectral type, according to Koo \etal (1996)
as verified by the DEEP collaboration (1999),
\begin{equation}
  \label{eq:nz}
  \frac{{\mathrm d}N}{{\mathrm d}z}(z) \simeq
   z^{2}e^{-\left(z/z_m\right)^2}~,
\end{equation}
\noindent where
\begin{equation}
  \label{eq:itoz}
   z_m = 0.722 + 0.149(I - 22.0)
\end{equation}
\noindent(Lanzetta, Yahil \& Fernandez-Soto 1996). SNAP colors were then
inferred by integrating the SED across filter profiles, adding an amount of
noise corresponding to the exposure time and instrument throughput.

{\sc Hyperz} was then used again, to estimate redshifts for the simulated
catalog as if it were real data. Unlike the image simulations in
$\S\ref{sims}$, this approach can already be taken to the depth of both the
wide and the deep SNAP surveys by extrapolating functional forms for the
luminosity and redshift distributions (eqs.~[\ref{eq:dNdI}] \& [\ref{eq:nz}]).
Magnitude cuts were applied at AB 26.5 (wide) or AB 29.1 (deep) in $R$. Similar
magnitude cuts were made in each filter, chosen at the $10\sigma$ detection
level of an exponential disc galaxy with FWHM=0.12$\arcsec$ (Kim \etal 2002). Past experience with lensing data (see $\S\ref{method}$, Bacon \etal 2000) confirms that this is reasonable $S/N$ limit. Note however that
the size and ellipticity cuts implemented for the simulated images in
$\S\ref{method}$ were not included at this stage.

\begin{figure}[htbp]
 \plotone{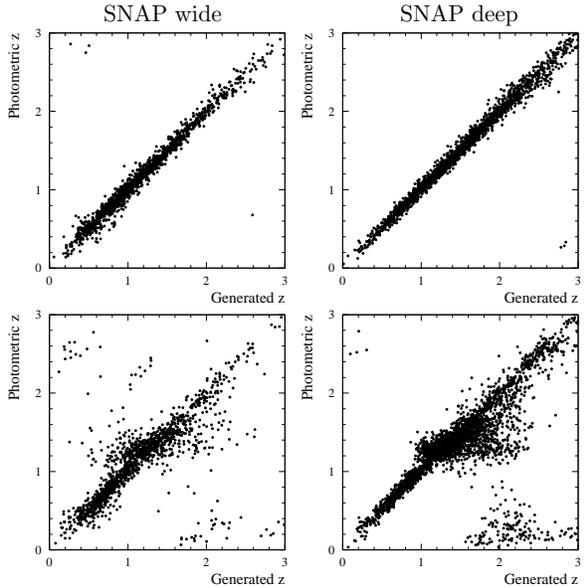}
 \caption{\small Recovery of redshifts of a realistic population of galaxies using
 {\sc hyperz} with the SNAP filter set. {\it Top-left panel}: the wide survey, 
 using all 9 colors. {\it Top-right panel}: the deep survey using all 9 colors. 
 {\it Bottom row}: the same, but with only the 6 optical colors, as if the 
 near IR HgCdTe data were not available.}
 \label{fig:zinout}
\end{figure}

\begin{figure}[htb]
 \plotone{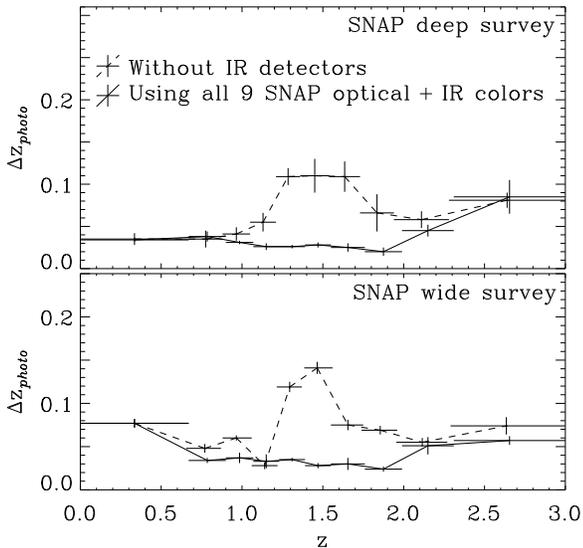}
 \caption{\small $\Delta z_{\rm photo}$, the rms scatter on photometric error
 estimation, as a function of increasing source redshift, $z$. {\it Top panel}:
 results for the SNAP deep survey. {\it Bottom
 panel}: results for the SNAP wide survey. In both cases, the solid line
 shows photometric redshift errors using observations in all 9 SNAP optical
 and near IR colors. The dashed lines show errors if the near IR
 HgCdTe data were unavailable. }
 \label{fig:dz-z}
\end{figure}

Figure $\ref{fig:zinout}$ shows the precision of photometric redshifts in both
the wide and deep SNAP surveys. All galaxy morphological types are included in
this analysis. Clearly demonstrated is the need for the near IR HgCdTe
detectors, a component of the satellite where a spacecraft has a clear
advantage over the ground.
Figure $\ref{fig:dz-z}$ shows the accuracy of the photo-$z$s as a function of
source (photometric) redshift. Here, $\Delta z_{\rm photo}(z)$ is the rms of
the core Gaussian in a double-Gaussian fit to horizontal slices through the
distributions in figure \ref{fig:zinout}.

To estimate the accuracy of 3D mass reconstructions (\S\ref{3dmap}), we now
concentrate on objects closer than $z=1$. According to equations
[\ref{eq:dNdI}] and [\ref{eq:nz}], these make up $\sim 38$\% of all galaxies
detected in $R$ for the wide SNAP survey, and $\sim 35$\% for the deep. For the
lensing analysis ($\S$\ref{method}), we have to reject some fraction of
galaxies because they were too small \& not resolved. Here, we assume that the
same percentage of rejection applies to the $z<1$ sub-sample of galaxies. This
will yield a conservative estimate of the number of objects remaining in the
real SNAP survey because objects closer than z=1 are likely to have a larger
median size than the entire sample. Removing this fraction from the number
density of galaxies shown in figure $\ref{fig:texp}$ leaves $40 \pm 5$ useful
galaxies per arcmin$^{2}$ in the SNAP wide survey to $z=1$, and more than
90arcmin$^{-2}$ in the SNAP deep survey. For these galaxies only, $\Delta
z_{deep}=$0.034 and $\Delta z_{wide}=$0.38 using all nine SNAP colors.

\newpage
\section{Dark Matter Mapping} \label{maps}

In this section, we describe the prospects of a space-based weak lensing survey
for mapping the 2D and 3D distribution of dark matter. Because of the high
number density of background galaxies resolved from space, this is one
application where a mission like SNAP will fare particularly better than
surveys from the ground.

\subsection{2D Maps} \label{2dmap}

To simulate observational data, we begin with shear maps created by raytracing
through N-body simulations from Jain, Seljak \& White (2000). We then add noise
to these idealized data, corresponding to the predicted levels for SNAP or
observing conditions at the currently most successful ground-based facilities.
In each case, we then attempt to recover the input projected mass distribution
by inverting the map of the shear into a map of the convergence $\kappa$.
Convergence is proportional to the projected mass along the line of sight, by a
factor depending on the geometrical distances between the observer, source and
lensed galaxies (see {\it e.g.}\ Bartelmann \& Schneider, 2001).

\begin{figure}[p]
 \epsscale{0.72}
 \plotone{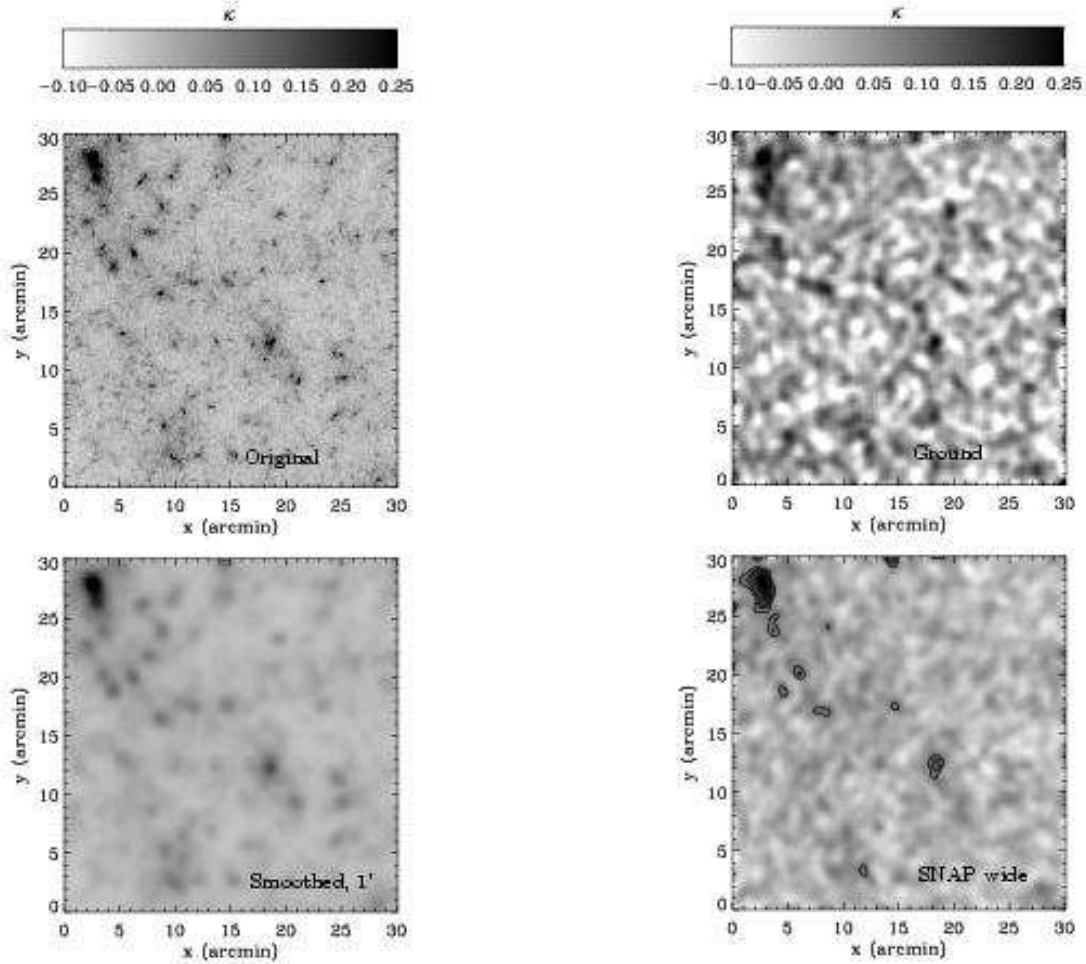}
 \epsscale{1}
 \caption{\small Left column. 2-dimensional convergence maps projected along a
 line of sight. The convergence $\kappa$ is proportional to the
 total matter density along the line of sight, and can be deduced from the
 shear field. {\it Top panel}: simulated (noise-free) convergence map derived by
 ray-tracing through an SCDM N-body simulations of large-scale structure
 from Jain, Seljak \& White (2000). The region shown is
 30$\arcmin\times$30$\arcmin$
 and the sources are assumed to lie at $z=1$. {\it Bottom panel}: same map but 
 smoothed using a Gaussian kernel with a FWHM of 1$\arcmin$.}
 \label{fig:maps}
 \caption{\small Right column. Reconstructions of the convergence map in
 figure \ref{fig:maps} which may be feasible from weak lensing surveys on
 the ground and from space. Overlaid contours show 3$\sigma$, 4$\sigma$ and  
 5$\sigma$ detection limits. {\it Top panel}: convergence $\kappa$ with noise 
 added corresponding to ground-based observations ({\it i.e.}\ $n_{g}=25$ 
 arcmin$^{-2}$ and $\sigma_{\gamma}=0.39$; Bacon \etal 2002).
 {\it Middle panel}: convergence map with the expected noise properties of the 
 wide SNAP survey ({\it i.e.}\ $n_{g}=105$ arcmin$^{-2}$ and
 $\sigma_{\gamma}=0.31$). {\it Bottom panel}: with the expected noise level of 
 the deep SNAP survey ({\it i.e.}\ $n_{g}=259$ arcmin$^{-2}$ and
 $\sigma_{\gamma}=0.36$).}
 \label{fig:maprecs}
\end{figure}

Figures~\ref{fig:maps} and \ref{fig:maprecs} show how the projected mass can be
mapped from space and from the ground. The color scale shows the convergence
$\kappa$. The top panel of figure~\ref{fig:maps} shows a (noise-free) simulated
convergence map from the ray tracing simulations of Jain, Seljak \& White
(2000) for an SCDM model. Underneath it is a version smoothed by a Gaussian
kernel with an rms of $1\arcmin$ for comparison to the simulated recovery from
observational data in figure~\ref{fig:maprecs}. 

Figure~\ref{fig:maprecs} shows similarly smoothed mass maps that would be
possible using (from top to bottom) a ground-based survey, the SNAP wide survey
and the SNAP deep survey. These were produced by adding to $\kappa$, before
smoothing, Gaussian random noise to each 1 arcmin$^{2}$ cell with an rms of
$\sigma_{\overline{\gamma}}$ given by equation~[\ref{eq:sigma_gamma}]. Overlaid
contours show mass concentrations detected at the 3$\sigma$, 4$\sigma$ and  
5$\sigma$ levels. For ground based observations, we set $\sigma_{\gamma}=0.39$,
and used $n_{g}=25$ arcmin$^{-2}$, as is available for ground-based surveys
({\it e.g.}\ Bacon \etal 2002). For the SNAP wide and deep survey, the surface
density of useable galaxies was taken to be $n_{g}=105$ arcmin$^{-2}$ and 259
arcmin$^{-2}$, and the rms shear noise per galaxy was taken to be
$\sigma_{\gamma}=$0.31 and 0.36, respectively, as derived from figure
\ref{fig:texp}. The galaxies are assumed to all have a redshift of $z$=1, which
is a good approximation as long as the median redshift is approximately that
value. As noted above, the surface density and median redshift will actually be
higher for the SNAP deep survey, because only exposure times corresponding to
that of the HDF were simulated.

\begin{figure}[p]
 \epsscale{0.72}
 \plotone{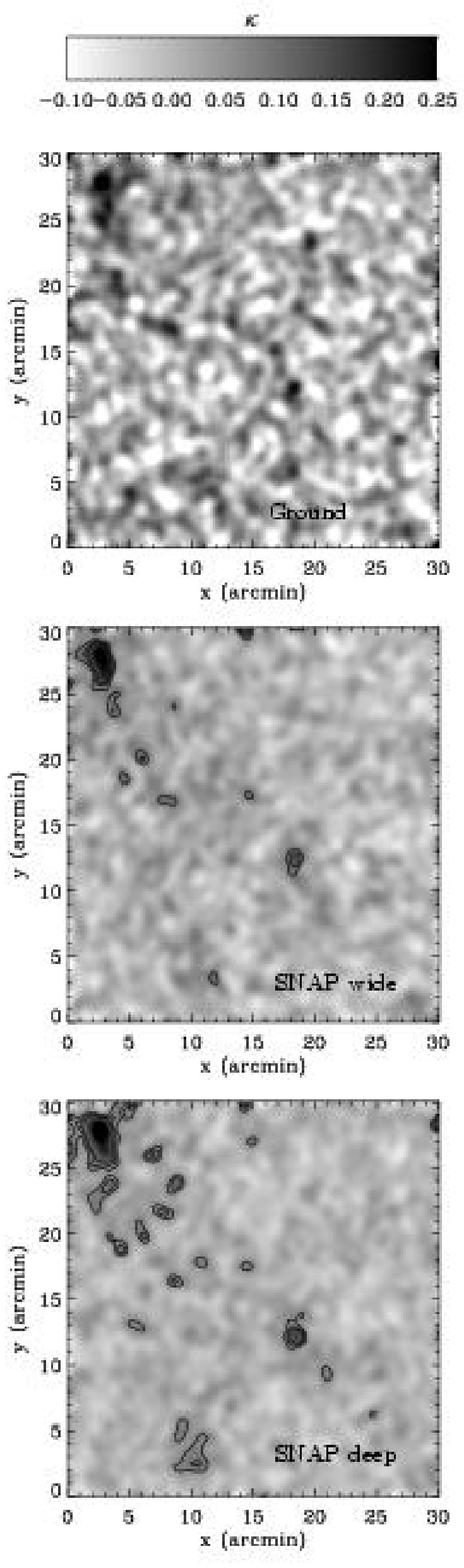}
 \epsscale{1}
 \vspace{2in}
\end{figure}

From the ground,only for the strongest features ({\it i.e.} the most massive
clusters) can a 3$\sigma$ detection be obtained. From space, the very high
density of resolved background galaxies allows the recovery of uniquely
detailed maps, including some of the filamentary structure and individual mass
overdensities down to the scale of galaxy groups and clusters. Thus, SNAP
offers the potential of mapping dark matter over very large fields of view,
with a precision well beyond that achievable with ground-based facilities.

The masses and locations of individual clusters can be extracted from such maps
using, for example, the $M_{\mathrm ap}$ statistic (Schneider 1996, 2002),
which has been applied successfully to find mass peaks in several surveys ({\it
e.g.}\ Hoekstra \etal 2002a; Erben \etal 2000) and our own work; or the
inversion method of Kaiser \& Squires (1993; KS) which was used by Miyazaki
\etal (2002). Marshall, Hobson \& Slosar (2003) demonstrate the effectiveness
of maximum entropy techniques to identify structures in KS lensing maps, using
criteria set by Bayesian evidence. White \etal (2002) argue that using any
detection method, a complete mass-selected cluster catalogue from 2D lensing
data would require a high rate of false-positive detections, since the prior
probability is for them to be {\emph anywhere} throughout a given survey. This
has been avoided in practice by secondary cross-checks of the lensing data with
spectroscopy, deep $x$-ray temperature or SZ observations. Indeed, two
previously unknown clusters have already been found in weak lensing maps and
spectroscopically confirmed by Wittman \etal (2001, 2003). However, this
confusion does make it harder to resolve the debate on the possible existence
of baryon-poor ``dark clusters'' ({\it e.g.}\ Dahle \etal 2003). These are a
speculative population of clusters which would be physically different to and
absent from the catalogues of optically or $x$-ray selected clusters. Remaining
dark-lens candidates (Erben \etal 2000; Umetsu \& Futamase 2000; Miralles \etal
2002) have currently been eliminated as chance alignments of background
galaxies (or possibly associations with nearby ordinary clusters Gray \etal
2001; Erben \etal 2003). If others could be found in high S/N weak lensing
maps, they would present a challenge to current models of structure formation,
and need to be accounted for in estimates of $\Omega_{\rm m}$; but they would
be unique laboratories to decipher the nature of dark matter.

\subsection{3D Maps} \label{3dmap}

The growth of mass structures can be followed in a rudimentary way via
photometric redshifts, by making 2D mass maps or power spectra with source
galaxies in different redshift slices (see {\it e.g.}\ Bartelmann \& Schneider
2000 \S4). This technique is useful for a global statistical analysis of a
survey in order to constrain cosmological parameters. It is used as such in
paper III, to predict possible constraints with SNAP. Tomographic measurements
of shear have also led to estimates of mass and radial position of clusters
(Wittman \etal 2001, 2003). After this analysis, spectroscopic redshifts were
needed to constrain the mass further by fixing the precise radial position of
clusters.

An alternative approach, in which one naturally reclaims the radial mass
information as well as the transverse density, has been developed by Taylor
(2003a) and Hu \& Keeton (2002). In this method, the shear pattern on an image
is treated as a fully 3D field, by including from the outset the redshift of
galaxy shear estimators as well as their 2D position on the sky. Taylor (2003a)
shows that there is a simple inversion that relates this 3D distortion field to
the underlying 3D gravitational potential.

Using this technique, we now demonstrate the capabilities of SNAP for
reconstructing the 3D mass distribution and locating clusters. We apply the
simulations of Bacon \& Taylor (2003) to the telescope and survey parameters
deduced in paper I, \S\ref{sims} and \S\ref{sens} then attempt to recover the
gravitational potential of two $M=10^{14}M_\odot$ NFW clusters at redshifts of
0.25 and 0.4 and separated by 0.2 degrees on the sky (see Bacon \& Taylor,
figures 4 \& 5). Note that this is specifically a search for clusters, which
induce a significant shear signal at one location, rather that integrating the
impact of many small objects and filamentary structures in a statistical basis
over an entire shear field. Our relatively simple input model is therefore
appropriate for our current purposes: it is a common occurrence that, for a
line of sight with large shear, a single cluster along a line of sight is
responsible for the signal.

First, we calculate the corresponding lensing potential for this field (using
the prescription of Bacon \& Taylor, equations 9-12; {\it c.f.}\ Kaiser \&
Squires 1993). As the lensing potential field $\phi$ is an integral of the
shear field, we are able to reconstruct $\phi$ with more accuracy than the
gravitational potential $\Phi$, which is a function of the second derivative of
$\phi$; nevertheless $\phi$ itself contains valuable 3D information. This is
discussed in full in Bacon \& Taylor (2003). We have used the expected
number density of useable galaxies closer than $z=1$ to be a conservative
$n_{g}(z<1)=90$ per arcmin$^{2}$, combining results from figure \ref{fig:texp}
and $\S\ref{photoz}$. In this nearby regime we can approximate
$\frac{dn}{dz}\propto z^2$. We have taken into account the shot noise arising
from intrinsic galaxy ellipticities, using an error on shear estimators for
galaxies of $\sigma_\gamma=0.36$. We have also included an error on our
photometric redshifts of $\Delta z_{\rm photo} = 0.034$ throughout $0<z<1$,
from $\S\ref{photoz}$.

\begin{figure}[htbp]
 \plotone{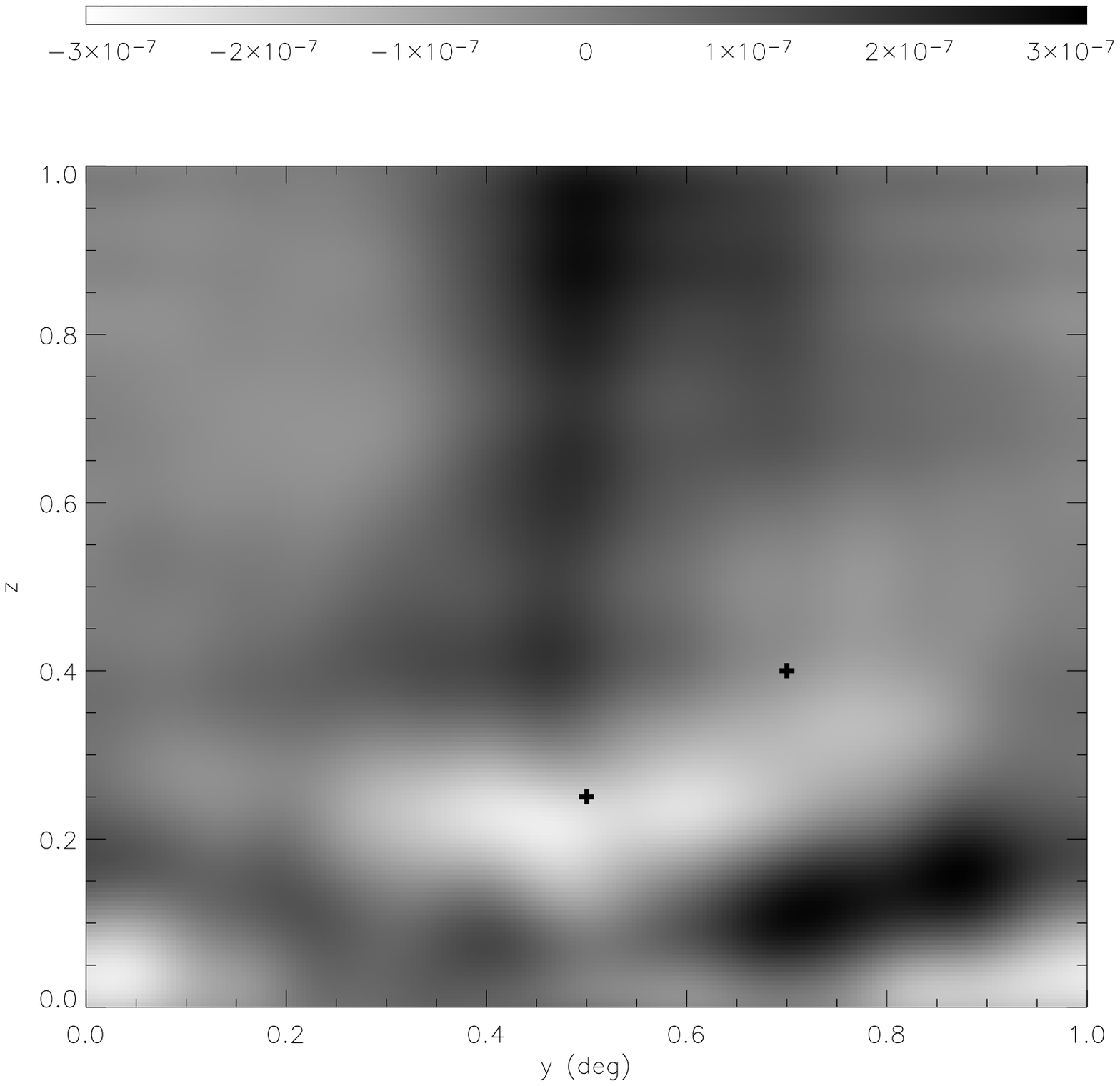} \\ 
 \plotone{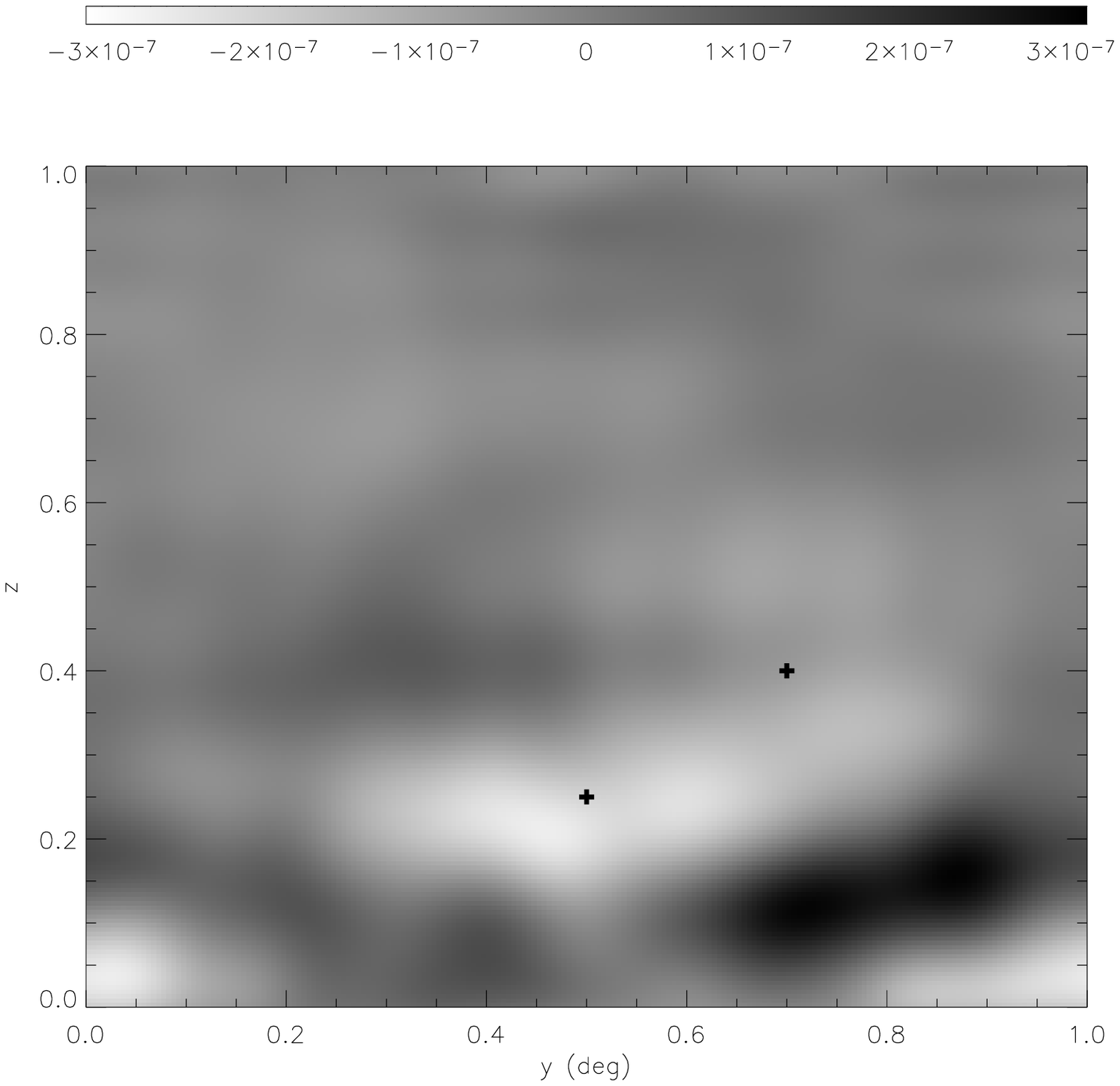}    
 \caption{\small {\it Top panel}: reconstructed lensing potential from a
 finite number of $z<1$ galaxies with realistic ellipticities; $n_{g}=90$
 arcmin$^{-2}$, $\sigma_\gamma = 0.36$, as expected for the SNAP
 deep survey. The $x$-axis represents angle in degrees; the $y$-axis
 represents redshift. The two crosses mark the positions of the input
 $M=10^{14}M_\odot$ NFW clusters. {\it Bottom panel}: difference between 
 input and recovered lensing potential fields.}
\label{fig:lensrec}
\end{figure}

Figure~\ref{fig:lensrec} shows the reconstruction of the lensing potential out
to $z=1$ available with the SNAP deep survey. The units of the lensing
potential here are radians$^2$, having chosen the differential in $\kappa=0.5
\partial^2 \phi$ to be taken in units of radians. In this simulation, we see
that the lower redshift cluster is very pronounced in the lensing potential,
with S/N per pixel of 5.4 at z=1. The lensing potential due to the higher
redshift cluster is also clearly visible, with S/N per pixel of 3.0.

\begin{figure}[htb]
 \plotone{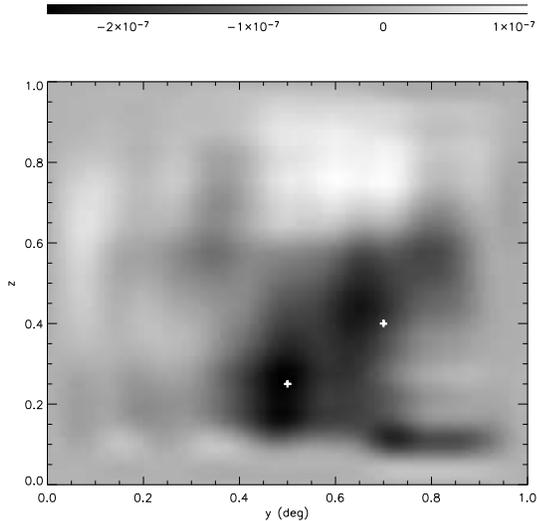}
 \caption{\small Reconstructed gravitational potential using galaxy properties of the
 SNAP deep survey, as in figure \ref{fig:lensrec}. The $x$-axis represents angle
 in degrees; the $y$-axis represents redshift. The two input clusters are
 clearly seen as the two darkest regions near the center of the image, marked
 with a cross.}
 \label{fig:gravrec}
\end{figure}

Figure~\ref{fig:gravrec} shows a reconstruction of the 3D gravitational
potential, using Taylor's inversion and Wiener filtering (Bacon \& Taylor
equations 8, 40). Even with a simulated mass of only $10^{14}M_\odot$, the
lower redshift cluster is very pronounced, and the higher redshift cluster is
also detectable at the 4.4$\sigma$ level. Extra noise peaks in
figure~\ref{fig:gravrec} demonstrate that the extremely low end of cluster
catalogues will be subject to high false detection rates. However, this
reconstruction affords measurement of masses of matter concentrations to an
accuracy of $\Delta M \simeq 1.1 \times 10^{13} M_\odot$ at $z\simeq 0.25$ or
$\Delta M \simeq 4.8 \times 10^{13} M_\odot$ at $z\simeq 0.5$ via $\chi^2$
fitting ({\it c.f.}\ Wittman \etal 2001, 2003). We can also estimate radial
position of mass concentrations from the simulated lensing data with accuracy
$\Delta z \simeq 0.05$ for clusters of mass $M = 10^{14} M_\odot$ at $z<0.5$
({\it c.f.}\ Bacon \& Taylor 2003). The mass concentrations are observed at
$z=0.25$ and 0.4 as expected with peak S/N of 2.8 and 3.3 respectively (N.B.
this is S/N per pixel; the overall detection significance of the cluster is as
quoted above). Of course, the sensitivity of this technique drops for clusters
at greater distances, as their induced lensing potential grows less within the
observed redshift window ({\it i.e.}\ $z<1$). In an alternative r\'{e}gime of
interest, mass fluctuations $\delta \sim 1$ are measurable on degree scales
({\it c.f.}\ Hu \& Keeton 2002).

Equivalent simulations can be carried out for ground-based experiments,
providing more limited prospects. The key difference that makes a space-based
experiment superior over a ground-based experiment in this regard is the
reduced error on shear estimates for galaxies, particularly for galaxies at
$z>0.5$, due to improved resolution and small PSF. From the ground, studies of
the 3-D $\phi$ field are restricted to measuring the mass of a cluster along
the line of sight at the $\Delta M \simeq 2 \times 10^{13} M_\odot$ level at
$z=0.25$, with a $1.3\sigma$ measurement of a mass of $10^{14} M_\odot$ at
$z=0.5$ along the line of sight of a foreground $z=0.25$ cluster.
Reconstruction of $\Phi$ in 3D is possible on 5$\arcmin$ scales only out to a
redshift of $z \simeq 0.5$ (see Bacon \& Taylor 2003). Application of the full
3D inversion technique to real ground-based data is currently being carried
out, and even measurements of one cluster behind another cluster are possible
(see review by Taylor 2003a).

SNAP's ability to measure the 3D gravitational potential in this fashion is of
great importance. One can determine the mass and density profile of several
matter concentrations along a line of sight, avoiding the ambiguity of
surface-density lensing or projection effects, and obtaining accurate
measurements of the mass of matter clumps in 3D. One can directly compare the
visible matter distribution with the underlying mass distribution to obtain
important information regarding biasing and galaxy formation as a function of
redshift. One can also examine the number of objects exceeding a certain mass
threshold as a function of redshift (see {\it e.g.}\ Viana \& Liddle 1996), or
reconstruct the 3D power spectrum directly (see Taylor 2003a) in order to
obtain constraints on cosmological parameters or to test the gravitational
instability paradigm which is thought to govern structure formation.

\section{Conclusions} \label{conc}

We have shown how a space-based wide field imager like SNAP is
ideally suited for studies of weak gravitational lensing. The aspects
of this satellite's design relevant for weak lensing, and the baseline
survey strategy, were presented in paper I. A shapelet-based method
for creating simulated space-based images (Massey \etal 2003,
Refregier 2003) has been used to predict SNAP's sensitivity to shear,
taking explicitly into account its instrumental throughput,
limitations and sensitivity.


In this paper, we have considered the baseline SNAP design for our
predictions. As explained in paper I, this design is almost optimal
because many requirements to find supernov\ae\ are the same as those to
measure weak lensing (stable imaging, small PSF, excellent multicolor
photometry).

The increased image resolution available from space makes possible the
construction of high resolution projected dark matter maps with an rms shear
sensitivity of $\sim$2.5\% in every 1$\arcmin$ cell for the 300 square degrees
wide SNAP survey and better than 1.8\% for the SNAP deep survey ({\it
c.f.}\ expected mean signal in a $\Lambda$CDM universe is approximately 3\%).
Since lensing is sensitive to mass regardless of its nature and state, these
maps will be unique tools for both astrophysics and cosmological parameter
estimation. Statistical properties of the dark matter distribution will be
precisely measured at several cosmological epochs and  constraints on
$\Omega_{\rm m}$, $\sigma_8$ and $w$ are discussed in paper III.

SNAP's simultaneous 9-band observations also open up new opportunities for
3-dimensional mapping via photometric redshift estimation (Taylor 2003a, Hu \&
Keeton 2002, Bacon \& Taylor 2002). SNAP's photometry allows an excellent
resolution of $\Delta z=0.034$ in redshift. Here we have shown that SNAP will
measure mass concentrations in full 3D with a 1$\sigma$ sensitivity of
approximately $1 \times 10^{13} M_\odot$ at $z\simeq 0.25$ and $\simeq 5 \times
10^{13} M_\odot$ at $z\simeq 0.5$. In this fashion it will be possible to
directly trace the non-linear growth of mass structures, testing with high
precision the gravitational instability theory.

Space-based wide-field imaging can be combined with weak gravitational
lensing to produce 2D and 3D mass-selected cluster catalogs down to
the scale of galaxy groups. Mass and light in the local universe can
be mapped out with exquisite precision, thus offering exciting
prospects for both astrophysics and cosmology.

\section*{Acknowledgments}

We thank the Raymond and Beverly Sackler fund for travel support. AR was
supported in Cambridge by a PPARC advanced fellowship. JR was supported by an
NRC/GSFC Research Associateship. We thank Alex Kim for his tireless efforts and
the well of information that is Mike Lampton. We are grateful for useful
discussions with Douglas Clowe, Andy Fruchter and George Smoot. Thanks also to
Mike Levi and the entire SNAP team for collaboration and ongoing work.

\end{document}